\documentclass{article}

\usepackage[numbers]{natbib}
\usepackage[preprint]{neurips_2026}

\usepackage[utf8]{inputenc} 
\usepackage[T1]{fontenc}    
\usepackage{hyperref}       
\usepackage{url}            
\usepackage{booktabs}       
\usepackage{multirow}   
\usepackage{amsfonts}       
\usepackage{nicefrac}       
\usepackage{microtype}      
\usepackage{xcolor}         

\usepackage{amsmath}
\usepackage{bbm}
\usepackage{subcaption} 
\usepackage{graphicx}
\usepackage{wrapfig}
\usepackage{textgreek}
\usepackage{algorithm}
\usepackage{algpseudocode}
\usepackage{float}
\usepackage{enumitem}
\usepackage{subcaption} 

\title{Ligand-Conditioned Discrete Diffusion for Protein Sequence–Structure Co-Design}

\author{
    \textbf{Chen Wei$^{1,2}$\thanks{Email: \texttt{chen\_wei@comp.nus.edu.sg}.}} \quad
  \textbf{Fanding Xu$^{3}$} \quad
  \textbf{Minghao Sun$^{2}$} \quad
  \textbf{Zhiyuan Liu$^{2}$} \quad
  \textbf{Lin Wang$^{4}$} \\
  \textbf{Tianrui Jia$^{2}$} \quad
  \textbf{Yihang Zhou$^{2}$\thanks{Corresponding authors. Emails: \texttt{yihangjoe@foxmail.com}, \texttt{zhang@nus.edu.sg}.}} \quad
  \textbf{Yang Zhang$^{2,4}$\footnotemark[2]} \\
  \\
  $^{1}$Xi'an University of Posts \& Telecommunications 
  $^{2}$National University of Singapore \\
  $^{3}$Xi'an Jiaotong University  \\
  $^{4}$Institute of Systems Medicine, Chinese Academy of Medical Sciences \\
}

\begin{document}

\maketitle

\vspace{-2em}
\begin{abstract}

Proteins perform their biological functions through three-dimensional structures encoded by amino acid sequences, and ligand-binding protein co-design requires models that generate sequence-structure compatible proteins under explicit ligand constraints. Although continuous diffusion and flow-based models have enabled ligand-aware protein design in coordinate or latent feature spaces, existing discrete diffusion protein language models mainly operate over sequence or structure tokens without direct small-molecule conditioning. We introduce \textbf{ProtLiD$^2$}, a \textbf{Prot}ein \textbf{L}igand-conditioned \textbf{D}iscrete \textbf{D}iffusion model for protein sequence-structure co-design. ProtLiD$^2$ jointly generates amino-acid sequence and discrete structure tokens while incorporating ligand chemical and geometric information through geometry-aware cross-attention. Trained on over one million ligand-protein complexes, ProtLiD$^2$ extends masked discrete diffusion from general sequence-structure generation to ligand-aware functional protein design. We further propose maximum confidence-margin guided ReMask decoding, an inference-time self-correction strategy that retains high-confidence predictions and remasks uncertain tokens for later refinement. Experimentally, ProtLiD$^2$ improves global fold confidence over Complexa in ligand-conditioned whole-protein design, increasing TM-score from $0.672$ to $0.802$ and pLDDT from $64.55$ to $73.00$. In ligand-binding pocket co-design, ProtLiD$^2$ reduces active-site BB-RMSD from $3.46/3.40$~\AA{} for FAIR/PocketGen to $1.97$~\AA{}, and improves ligand-aware combined pass rates over PocketGen from $14.86\%$ to $59.73\%$ and from $6.08\%$ to $23.49\%$ under increasingly stringent docking thresholds. These results demonstrate the potential of ligand-conditioned discrete diffusion as an effective token-space framework for functional protein co-design. To promote further progress in ligand-aware protein design and enable rapid adoption in practical applications, the inference code will be made publicly available at \url{https://github.com/auroua/ProtLiD}.

\end{abstract}

\section{Introduction}

Proteins are fundamental biomolecules that fold from linear amino acid sequences into three-dimensional structures, enabling diverse functions that drive nearly every biological process across all forms of life, from catalysis and signaling to molecular recognition and cellular regulation. Due to their superior ability to learn the underlying distributions of large-scale training data, recent data-driven generative models have transformed protein design, shifting the field from traditional physics-based and evolutionary-profile-guided design methods \cite{KORTEMME200491, pearceEvoDesignDesigningProtein2019} toward diffusion-based generative models \cite{watsonNovoDesignProtein2023, ahernAtomlevelEnzymeActive2026, Butcher2025.09.18.676967, DBLP:conf/nips/ZhangLHZ023, zhangEfficientGenerationProtein2024, geffnerLaProteinaAtomisticProtein2025, didi2026scaling, Alamdari2023.09.11.556673, conf/icml/WangZYXHG24, conf/iclr/WangZYXHG25, DBLP:conf/icml/HsiehWZXYHZG25}. These generative approaches can be broadly categorized into continuous diffusion models \cite{watsonNovoDesignProtein2023, ahernAtomlevelEnzymeActive2026, Butcher2025.09.18.676967, zhangEfficientGenerationProtein2024, DBLP:conf/nips/ZhangLHZ023, geffnerLaProteinaAtomisticProtein2025, didi2026scaling} and discrete diffusion models \cite{Alamdari2023.09.11.556673, conf/icml/WangZYXHG24, conf/iclr/WangZYXHG25, DBLP:conf/icml/HsiehWZXYHZG25} according to their generation space: continuous diffusion models perform denoising over continuous coordinate or feature representations, whereas discrete diffusion models generate proteins through iterative refinement in tokenized sequence, structure, or joint sequence-structure spaces.

Continuous diffusion- and flow-based protein design models \cite{watsonNovoDesignProtein2023, ahernAtomlevelEnzymeActive2026, Butcher2025.09.18.676967, zhangEfficientGenerationProtein2024, DBLP:conf/nips/ZhangLHZ023, geffnerLaProteinaAtomisticProtein2025, didi2026scaling} have emerged as a powerful class of generative approaches that progressively denoise protein representations to design structures, ranging from backbone-level models to fully atomistic frameworks capable of ligand- or protein-conditioned sequence-structure co-design. In parallel with continuous diffusion models, discrete diffusion models \cite{Alamdari2023.09.11.556673, conf/icml/WangZYXHG24, conf/iclr/WangZYXHG25, DBLP:conf/icml/HsiehWZXYHZG25} have recently emerged as a complementary protein design paradigm, operating directly in amino-acid or tokenized structure space to generate, inpaint, and co-design protein sequences and structures through iterative denoising. \textbf{Although discrete diffusion protein language models have enabled unconditional sequence generation, motif scaffolding, inverse folding, folding, and sequence–structure co-generation, they still lack the explicit ligand-conditioning capability of continuous diffusion models, which can directly generate protein sequences and structures in the context of ligand constraints}. To fill this gap, we present \textbf{ProtLiD$^2$}, a \textbf{Prot}ein \textbf{Li}gand-conditioned \textbf{D}iscrete \textbf{D}iffusion model for protein sequence--structure co-design that jointly designs protein sequence and structure in discrete token space under explicit ligand conditioning, extending discrete diffusion protein design toward functional ligand-aware generation.

Existing discrete diffusion protein language models, including EvoDiff \cite{Alamdari2023.09.11.556673}, DPLM \cite{conf/icml/WangZYXHG24}, DPLM-2 \cite{conf/iclr/WangZYXHG25} and Geo-DPLM \cite{DBLP:conf/icml/HsiehWZXYHZG25}, generate proteins through iterative denoising in discrete sequence or structure-token space, typically unmasking multiple tokens in parallel via order-agnostic or confidence-ranked mask decoding. Recent advances in masked discrete diffusion language models \cite{wang2026remasking, arriola2025block, nie2026large, DBLP:conf/icml/KimSKKC25} suggest that the sampling trajectory, especially the token unmasking order, plays a critical role in generation quality. In this sense, token unmasking order provides a natural form of test-time scaling for masked discrete diffusion models, where additional inference-time computation is used to plan or refine the denoising trajectory rather than retrain the model. Motivated by this observation, we propose a maximum confidence-margin guided ReMask decoding strategy that retains high-certainty token predictions while remasking ambiguous ones for later refinement. This provides a lightweight inference-time self-correction mechanism that improves decoding stability and sequence--structure consistency without retraining or architectural modification.

In summary, we highlight our main contributions as follows:

\begin{enumerate}[label=(\roman*)]


    \item We propose ProtLiD$^2$, a ligand-conditioned masked discrete diffusion model for joint protein sequence-structure co-design. ProtLiD$^2$ represents proteins in a unified sequence-structure token space and incorporates ligand chemical and geometric information through geometry-aware cross-attention, extending discrete diffusion protein language modeling from general sequence-structure generation to ligand-aware functional protein design.


    \item We curate a large-scale ligand-protein complex dataset for ligand-conditioned sequence-structure co-design. After source-specific filtering and leakage removal against the PLINDER benchmark, the final training set contains 1 million ligand-protein complexes.
    
    \item We introduce a maximum confidence-margin guided ReMask decoding strategy for discrete diffusion sampling. This strategy preserves the stochastic reverse transition of MDLM while using confidence-margin scores to retain reliable token predictions and remask uncertain positions for later refinement, enabling inference-time self-correction that improves sampling stability and sequence--structure consistency without retraining or architectural modification.
    
    \item We systematically evaluate ProtLiD$^2$ across sequence-structure, ligand-conditioned whole-protein, and pocket co-design benchmarks, showing improved whole-protein fold confidence over Complexa and substantially better pocket-design accuracy than FAIR and PocketGen, including lower active-site BB-RMSD and higher ligand-aware combined pass rates.

    
\end{enumerate}


\section{Related Work}

\subsection{Generative Protein Design Model}

Continuous diffusion and flow-based models have advanced protein design from backbone generation followed by inverse folding \cite{doi:10.1126/science.add2187} to atomistic, ligand-aware, and conditioned generation. RFdiffusion-family models \cite{watsonNovoDesignProtein2023,ahernAtomlevelEnzymeActive2026,Butcher2025.09.18.676967} extend motif and binder scaffolding to atom-level functional conditioning, while pocket-design models such as FAIR \cite{DBLP:conf/nips/ZhangLHZ023} and PocketGen \cite{zhangEfficientGenerationProtein2024} jointly generate ligand-conditioned pocket sequences and atomic structures using refinement or atom/residue/ligand interaction modeling. Partially latent flow-matching models, including La-Proteina \cite{geffnerLaProteinaAtomisticProtein2025} and Proteina-Complexa \cite{didi2026scaling}, further enable joint sequence-structure and target-conditioned binder design in continuous latent spaces.

Discrete diffusion protein language models provide a complementary paradigm to continuous diffusion by generating proteins in amino-acid or tokenized structure spaces. EvoDiff \cite{Alamdari2023.09.11.556673} and DPLM \cite{conf/icml/WangZYXHG24} primarily target sequence generation and controllable tasks such as inpainting, motif scaffolding, and inverse folding, while DPLM-2 \cite{conf/iclr/WangZYXHG25} extends discrete diffusion to sequence-structure co-design through learned backbone tokenization and decoding \cite{DBLP:conf/iclr/JingESTD21,DBLP:conf/iclr/YuLGVSMCGGHG0ER24,jumperHighlyAccurateProtein2021a}. Geo-DPLM \cite{DBLP:conf/icml/HsiehWZXYHZG25} further improves structure-token modeling with enhanced supervision, refinement, and geometry-aware modules. Despite these advances, explicit ligand-conditioned co-design remains largely unexplored in discrete sequence-structure token space.


To fill this gap, rather than optimizing the structure tokenization module itself, this work adopts an existing protein structure tokenization model, GCP-VQVAE \cite{Pourmirzaei2025.10.01.679833}, and focuses on demonstrating that explicit ligand conditioning can be effectively integrated into discrete diffusion models for high-performing ligand-aware protein sequence-structure co-design, providing a complementary token-space alternative to continuous diffusion-based protein design. 

For ligand representation, recent studies have developed powerful molecular embedding models that encode chemical identity, atomic context, and molecular geometry \cite{DBLP:conf/iclr/ZhouGDZXWZK23, https://doi.org/10.1002/advs.202513556, liu2023molca,liu2025nextmol,luo2025towards,liu2024prott}. Following this line of work, We use Uni-Mol~\cite{DBLP:conf/iclr/ZhouGDZXWZK23} as the ligand encoder to extract contextual chemical and geometric embeddings from ligand atom types and 3D coordinates, allowing small-molecule information to condition the discrete diffusion process through cross-attention.

\subsection{Masked Discrete Diffusion Language Models}

Discrete diffusion protein language models typically use absorbing-state corruption, where clean tokens are progressively replaced by \texttt{[MASK]}. The DPLM series \cite{conf/icml/WangZYXHG24,conf/iclr/WangZYXHG25,DBLP:conf/icml/HsiehWZXYHZG25} adopts a reparameterized masked diffusion view \cite{zheng2024a}, treating generation as a route-and-denoise process that separates clean-token prediction from token selection.

In this work, we follow the masked discrete diffusion language model (MDLM) formulation \cite{DBLP:conf/nips/SahooASGMCRK24,DBLP:conf/nips/ShiHWDT24}. Given a clean token sequence $x_0$, the forward process independently preserves each token with probability $\alpha_t$ and replaces it with the mask token $m$ with probability \(q(x_t \mid x_0)=\prod_i \mathrm{Cat}\!\left(x_t^{(i)};\alpha_t x_0^{(i)}+(1-\alpha_t)m\right)\), 
where $\alpha_t$ is a decreasing masking schedule. The denoising model $\mu_\theta(x_t,t)$ predicts the clean-token distribution at masked positions, and the training objective reduces to a weighted masked cross-entropy loss:
\begin{equation}
\mathcal{L}
=
\int_0^1
w(t)\,
\mathbb{E}_{q(x_t \mid x_0)}
\left[
\sum_{i:x_t^{(i)}=m}
-\log \mu_\theta^{(i)}(x_t,t)_{x_0^{(i)}}
\right] dt,
\label{eq:mdlm-loss}
\end{equation}
where $w(t)=-\alpha_t'/(1-\alpha_t)$.

Generation reverses the masking process from an all-masked or partially masked sequence. At each reverse step from $t$ to $s<t$, unmasked tokens are copied unchanged, while each masked position is sampled as
\begin{equation}
x_s^{(i)} \sim 
\mathrm{Cat}
\left(
\frac{\alpha_s-\alpha_t}{1-\alpha_t}
\mu_\theta^{(i)}(x_t,t)
+
\frac{1-\alpha_s}{1-\alpha_t}
e_m
\right),
\quad \text{if } x_t^{(i)} = m,
\label{eq:mdlm-reverse}
\end{equation}
where $e_m$ denotes the one-hot mask token. This transition either reveals a token according to the denoising distribution or keeps it masked for later refinement.

\subsection{Unmasking Strategies}

In masked discrete diffusion models, token reveal order is a key factor in generation quality. While vanilla MDLM sampling unmasks tokens in an order-agnostic reverse process \cite{DBLP:conf/nips/SahooASGMCRK24,DBLP:conf/nips/ShiHWDT24}, adaptive strategies improve decoding by prioritizing high-confidence positions, such as those with large top-$K$ confidence or top-1/top-2 probability margins \cite{DBLP:conf/icml/KimSKKC25}. LLaDA \cite{nie2026large} and ReMDM \cite{wang2026remasking} further introduce low-confidence remasking to enable iterative refinement and inference-time scaling. Unlike prior adaptive unmasking methods, we propose Max Confidence-Margin ReMask, which decouples candidate proposal from token retention during decoding: candidate updates are first sampled from the MDLM reverse transition, after which high-margin predictions are retained and uncertain tokens are remasked for later refinement.

\section{Method}\label{method}

\begin{figure*}[t]
    \centering
    \includegraphics[width=\textwidth]{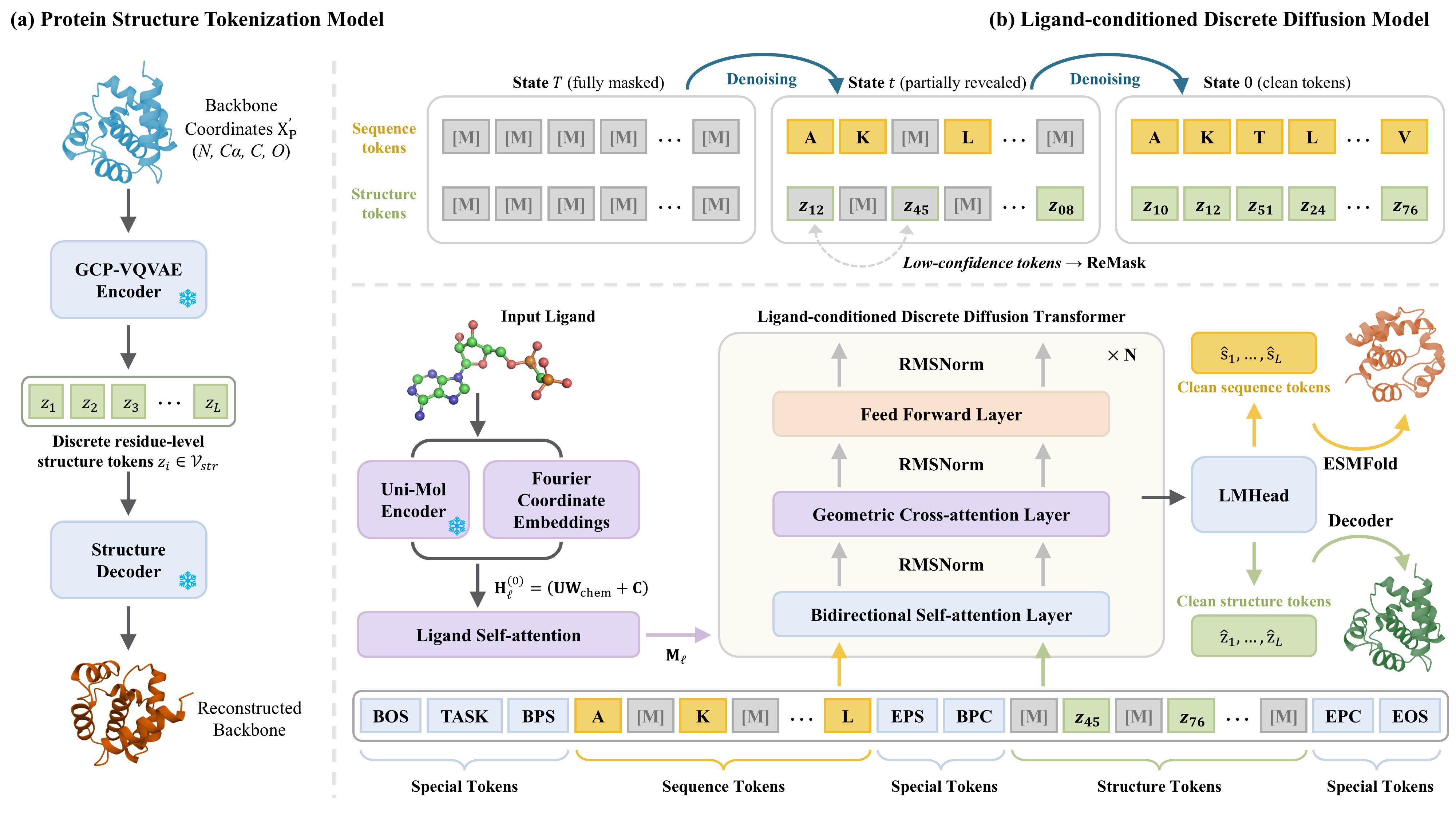}
\caption{
Overview of the proposed ProtLiD$^2$ model. 
(a) A frozen GCP-VQVAE tokenizer converts protein backbone coordinates into residue-level structure tokens. 
(b) ProtLiD$^2$ jointly denoises sequence and structure tokens with a ligand-conditioned masked discrete diffusion Transformer. 
Ligand chemical and geometric features are injected through geometry-aware cross-attention, and MCM-ReMask retains confident predictions while remasking uncertain tokens for refinement.
}
    \label{fig:architecture_overview}
\end{figure*}

An overview of the proposed architecture is shown in Fig.~\ref{fig:architecture_overview}. 
ProtLiD$^2$ represents proteins as paired amino-acid sequence tokens and discrete structure tokens obtained from a frozen GCP-VQVAE tokenizer. Ligands are encoded with atom-level chemical features and Fourier coordinate embeddings, and their information is injected into a masked discrete diffusion Transformer through geometry-aware cross-attention. During inference, the model iteratively denoises corrupted sequence and structure tokens, while the proposed MCM-ReMask strategy retains confident predictions and remasks uncertain positions to improve sampling stability and sequence-structure consistency.


\subsection{Dataset Construction} \label{datac}

To train ProtLiD$^2$, we constructed a large-scale ligand-conditioned protein sequence--structure dataset by integrating protein--ligand complexes from Protenix~\cite{2026.02.05.703733}, PLINDER~\cite{anonymous2024plinder}, CrossDock~\cite{doi:10.1021/acs.jcim.0c00411}, HiQBind~\cite{D4DD00357H}, and AlphaFill-derived complexes~\cite{Hekkelman2023}. Complexes from non-Protenix sources were processed with the AlphaFold3 data-processing pipeline~\cite{Abramson2024} to obtain a unified representation of protein chains, ligand identities, ligand coordinates, and protein backbone geometry. 

For each complex, we extracted one protein chain and its associated ligand, retaining examples with valid protein coordinates, ligand SMILES, and at least one ligand-contacting residue within a 6.0~\AA{} cutoff. We further removed long proteins, large ligands, severe protein--ligand clashes, and low-confidence AlphaFill-derived complexes. After filtering and source-specific sampling, the merged dataset contained 1,125,038 ligand-protein complexes. To reduce redundancy, samples were indexed by protein sequence so that each unique sequence could be associated with one or more complexes. Detailed filtering criteria and source-specific statistics are provided in Appendix~\ref{app:dataset_details}.


For leakage-aware evaluation, we used the PLINDER test set and removed training examples with $\geq 30\%$ sequence identity to any PLINDER test protein using MMseqs2~\cite{Steinegger2017}, yielding 1,026,766 training complexes. Due to the cost of full test set evaluation, we sampled 200 protein-ligand complexes with an approximately uniform sequence-length distribution as the final benchmark set.

\subsection{Ligand-Conditioned Sequence-Structure Co-design Model}



ProtLiD$^2$ formulates ligand-conditioned protein design as masked discrete diffusion over a unified sequence-structure token space. Given a protein-ligand complex, the protein is represented by paired amino-acid sequence and discrete structure tokens, while the ligand is encoded from atom-level chemical features and 3D coordinates. The model learns to recover clean protein tokens from \texttt{[MASK]}-corrupted inputs under ligand conditioning.

Before diffusion denoising, ProtLiD$^2$ preprocesses each protein-ligand complex into two inputs: a unified protein sequence-structure token sequence and a ligand conditioning representation derived from chemical and geometric features. \textbf{Protein and ligand preprocessing.} For each protein-ligand complex, the protein is represented by its amino-acid sequence $\mathbf{a}=(a_1,\ldots,a_L)$ and backbone coordinates $\mathbf{X}_{\mathrm{p}}\in\mathbb{R}^{L\times4\times3}$, while the ligand is represented by heavy-atom coordinates $\mathbf{X}_{\ell}\in\mathbb{R}^{M\times3}$ and atom-level features. To reduce global SE(3) variation while preserving protein-ligand geometry, we canonicalize each complex by centering it at the protein C$\alpha$ centroid and applying a ligand-guided PCA rotation to both protein and ligand coordinates. Further details are provided in Appendix~\ref{app:model_details}. \textbf{Protein sequence and structure tokenization.} The amino-acid sequence is tokenized as $\mathbf{s}=(s_1,\ldots,s_L)$, where $s_i\in\mathcal{V}_{\mathrm{seq}}$. As illustrated in Fig.~\ref{fig:architecture_overview}(a), a frozen GCP-VQVAE tokenizer maps the canonicalized backbone coordinates to residue-level discrete structure tokens $\mathbf{z}=(z_1,\ldots,z_L)$, where $z_i\in\mathcal{V}_{\mathrm{str}}$. Sequence and structure tokens are concatenated into a unified multimodal token sequence,
$\mathbf{y}_0=[\mathtt{BOS},\mathtt{TASK},\mathtt{BPS},s_1,\ldots,s_L,\mathtt{EPS},\mathtt{BPC},z_1,\ldots,z_L,\mathtt{EPC},\mathtt{EOS}]$,
where special tokens mark the sequence and structure spans. Structure-token indices are shifted so that both modalities are represented in a shared vocabulary while retaining modality-specific validity constraints. \textbf{Ligand representation.} As shown in Fig.~\ref{fig:architecture_overview}(b), the ligand condition is encoded from both chemical and geometric information. We use Uni-Mol~\cite{DBLP:conf/iclr/ZhouGDZXWZK23} to extract atom-level and pairwise atom--atom features, and encode ligand coordinates using Fourier coordinate embeddings. The projected chemical features and coordinate embeddings are fused and refined by stacked pairwise-aware ligand self-attention layers, producing the final ligand memory $\mathbf{M}_{\ell}$. Details of the ligand embedding and pair-bias attention are given in Appendix~\ref{app:model_details}. \textbf{Geometry-aware ligand cross-attention.} To inject ligand information into the protein denoising backbone, we insert a geometry-aware ligand cross-attention layer after each protein self-attention block. Protein hidden states are used as queries, while the ligand memory $\mathbf{M}_{\ell}$ provides keys and values. In addition to standard attention logits, we add a learned geometric bias derived from pairwise distances between protein-token proxy coordinates and ligand atom coordinates. This allows sequence and structure tokens to attend to chemically encoded ligand atoms while emphasizing spatially relevant protein--ligand interactions. The full formulation is provided in Appendix~\ref{app:model_details}.

\paragraph{Ligand-conditioned discrete diffusion transformer.}
The denoising backbone follows the architecture shown in Fig.~\ref{fig:architecture_overview}(b). At diffusion time $t$, the clean sequence-structure token sequence $\mathbf{y}_0$ is corrupted into $\mathbf{y}_t$ by an absorbing-state masking process. The initial hidden states are obtained from the noisy tokens as $\mathbf{H}^{(0)}=\mathrm{Embed}(\mathbf{y}_t)$. These hidden states are processed by stacked Transformer blocks with bidirectional self-attention, ligand-conditioned geometric cross-attention, and feed-forward layers. After $N$ layers, the final hidden states are passed through two fully connected layers, denoted as $\mathrm{LMHead}$, to produce logits over the joint sequence-structure vocabulary:
$\boldsymbol{\ell}_{\theta}(\mathbf{y}_t,\mathbf{M}_{\ell})=\mathrm{LMHead}(\mathbf{H}^{(N)})$.
The corresponding denoising distribution is $\mu_{\theta}(\mathbf{y}_t,\mathbf{M}_{\ell})=\mathrm{Softmax}(\boldsymbol{\ell}_{\theta}(\mathbf{y}_t,\mathbf{M}_{\ell}))$, which predicts the clean sequence and structure tokens from the corrupted input.

Following the MDLM objective in Eq.~(1), the model is trained with weighted masked cross-entropy over the joint sequence-structure token sequence. During training, only masked non-special tokens contribute to the loss, and modality-specific vocabulary constraints are used so that sequence positions are predicted over $\mathcal{V}_{\mathrm{seq}}$ and structure-token positions are predicted over $\mathcal{V}_{\mathrm{str}}$.  During generation, tokens are progressively sampled from the reverse transition defined in Eq.~(2), which either reveals a token according to the model predicted denoising distribution or keeps the position masked for later refinement.

\subsection{Maximum Confidence-Margin Guided ReMask Decoding Strategy}

To improve the robustness of discrete diffusion sampling, we introduce MCM-ReMask, a maximum confidence-margin guided ReMask decoding strategy. At each reverse step, candidate token updates are first proposed by the original MDLM reverse transition in Eq.~(2), thereby preserving the stochastic reveal process of masked diffusion. We then verify each proposed token using the probability margin between the top-1 and top-2 predictions computed from the model logits: high-margin candidates are retained as reliable updates, whereas ambiguous or invalid candidates are returned to \texttt{[MASK]} for later refinement. This verification-and-remasking procedure provides a lightweight inference-time self-correction mechanism that stabilizes the sampling trajectory and improves sequence-structure consistency.

\begin{algorithm}[!t]
\caption{Max Confidence-Margin Guided ReMask Decoding}
\label{alg:max_margin_remask}
\begin{algorithmic}[1]
\Require Current sequence $x_t$, MDLM reverse transition $q(x_s\mid x_t,\hat{x}_0)$ defined by Eq.~(2), model logits $\ell$, mask token $\mathtt{[MASK]}$
\Ensure Updated sequence $x_s$

\State $M_{\mathrm{active}} \gets (x_t=\mathtt{[MASK]})$, \quad $x_s \gets x_t$

\For{each $i$ with $M_{\mathrm{active}}^{(i)}=\mathrm{True}$}
    \State Sample $x_s^{(i)} \sim q(x_s^{(i)}\mid x_t,\hat{x}_0)$ according to Eq.~(2)
\EndFor

\State $k \gets \sum_i \mathbf{1}\!\left[
x_t^{(i)}=\mathtt{[MASK]}
\land
x_s^{(i)}\neq\mathtt{[MASK]}
\right]$


\State \textbf{if } $k \le 0$ \textbf{ then } set $x_s^{(i)}\gets\mathtt{[MASK]}$ for all $i$ with $M_{\mathrm{active}}^{(i)}$ and \Return $x_s$

\State Construct constrained logits $\tilde{\ell}$ by suppressing $\mathtt{[MASK]}$ and invalid token types



\For{each position $i$}
    \State $c^{(i)} \gets 
    \begin{cases}
    \left|p_1^{(i)}-p_2^{(i)}\right|, & M_{\mathrm{active}}^{(i)}=\mathrm{True},\\
    -\infty, & \text{otherwise},
    \end{cases}$
    \Statex \hspace{\algorithmicindent}where $(p_1^{(i)},p_2^{(i)})=\operatorname{TopK}(\operatorname{Softmax}(\tilde{\ell}^{(i)}),2)$
\EndFor

\State $k \gets \min(k, |\{i:c^{(i)}\neq-\infty\}|)$

\State $S \gets \operatorname{TopK}(c,k)$ \Comment{positions whose sampled candidates are retained}

\While{$\exists i \in S$ such that $x_s^{(i)}=\mathtt{[MASK]}$}
    \State Set $c^{(i)} \gets -\infty$ for all $i\in S$ with $x_s^{(i)}=\mathtt{[MASK]}$
    \If{no valid candidates remain}
        \State \Return $x_s$
    \EndIf
    \State $k \gets \min(k, |\{i:c^{(i)}\neq-\infty\}|)$
    \State $S \gets \operatorname{TopK}(c,k)$
\EndWhile


\For{each $i$ with $M_{\mathrm{active}}^{(i)}=\mathrm{True}$}
    \State $x_s^{(i)} \gets 
    \begin{cases}
    x_s^{(i)}, & i\in S \quad \text{retain sampled candidate},\\
    \mathtt{[MASK]}, & i\notin S \quad \text{remask uncertain position}.
    \end{cases}$
\EndFor

\State \Return $x_s$

\end{algorithmic}
\end{algorithm}

\section{Experiment} \label{exp}


We evaluate ProtLiD$^2$ on three complementary protein design settings: unmasking-strategy evaluation, ligand-conditioned whole-protein co-design, and ligand-binding pocket co-design. The first setting examines whether the proposed MCM-ReMask decoding strategy improves sequence-structure self-consistency across a wide range of protein lengths. The ligand-conditioned whole-protein setting evaluates whether the model can generate globally plausible protein structures under ligand constraints. Finally, the pocket co-design setting focuses on the most practically relevant local design problem, where the model must preserve both the global fold and the ligand-binding microenvironment.




\subsection{Experimental Setup} \label{setup}

ProtLiD$^2$ is implemented as a Transformer-based model \cite{DBLP:conf/nips/VaswaniSPUJGKP17} with approximately 370M parameters. It contains 16 Transformer layers with hidden dimension 1280, feed-forward dimension 5120, and 10 attention heads. The model was trained with PyTorch on 8 NVIDIA A6000 GPUs, each with 96GB memory, for 100,000 optimization steps over approximately 11 days. Training used bfloat16 precision, gradient accumulation over 8 steps, and with a maximum of 45,000 tokens per batch and maximum sequence length 1024. We optimized the model with AdamW using learning rate $6 \times 10^{-4}$, $\beta=(0.9, 0.95)$, weight decay 0.1, gradient clipping at 1.0, 10,000 warmup steps, and a cosine learning-rate schedule. Geometric data augmentation was applied using random rotations with probability 0.3 and coordinate noise with scale 0.07.

For unmasking-strategy evaluation, we compare MCM-ReMask with representative decoding strategies, including ReMDM~\cite{wang2026remasking}, LLaDA-ReMask~\cite{nie2026large}, and TopK-Margin~\cite{DBLP:conf/icml/KimSKKC25}. For each strategy, we generate 100 protein sequence-structure pairs at each target length from 100 to 700 residues. Because ProtLiD$^2$ is ligand-conditioned, each generation uses a randomly sampled ligand to preserve the model's conditioning architecture. Sequence-structure self-consistency is evaluated by comparing the GCP-VQVAE-decoded backbone from generated structure tokens with the ESMFold~\cite{doi:10.1126/science.ade2574} predicted structure from the generated sequence, using TM-score~\cite{https://doi.org/10.1002/prot.20264}, RMSD~\cite{MAIOROV1994625} of backbone atoms (BB-RMSD), and pLDDT\cite{jumperHighlyAccurateProtein2021a}.


For ligand-conditioned whole-protein and pocket co-design evaluations, each method generates 10 candidates per target under the same ligand condition. Whole-protein co-design additionally conditions on the target protein length, whereas pocket co-design fixes the non-pocket context and redesigns ligand-contacting residues within a 6.0~\AA{} protein-ligand heavy-atom cutoff. We evaluate sequence-structure self-consistency by folding generated sequences with ESMFold and comparing them with model-generated structures using BB/CA-RMSD, TM-score, and pLDDT. For ligand-aware evaluation, we select the highest-pLDDT candidate per target, predict the protein-ligand complex with AlphaFold3, and compute an AF3-Vina score using AutoDock Vina~\cite{doi:10.1021/acs.jcim.1c00203} with a ligand-centered docking box. We also report combined pass-rate criteria integrating fold confidence, structural accuracy, and ligand-aware docking quality, with full definitions in Appendix~\ref{app:whole_protein_eval} and ~\ref{app:pocket_eval}.

\subsection{Unmasking Strategy Evaluation for Protein Co-design}

\begin{figure*}[t]
\centering

\begin{subfigure}{0.32\textwidth}
    \centering
    \includegraphics[width=\linewidth]{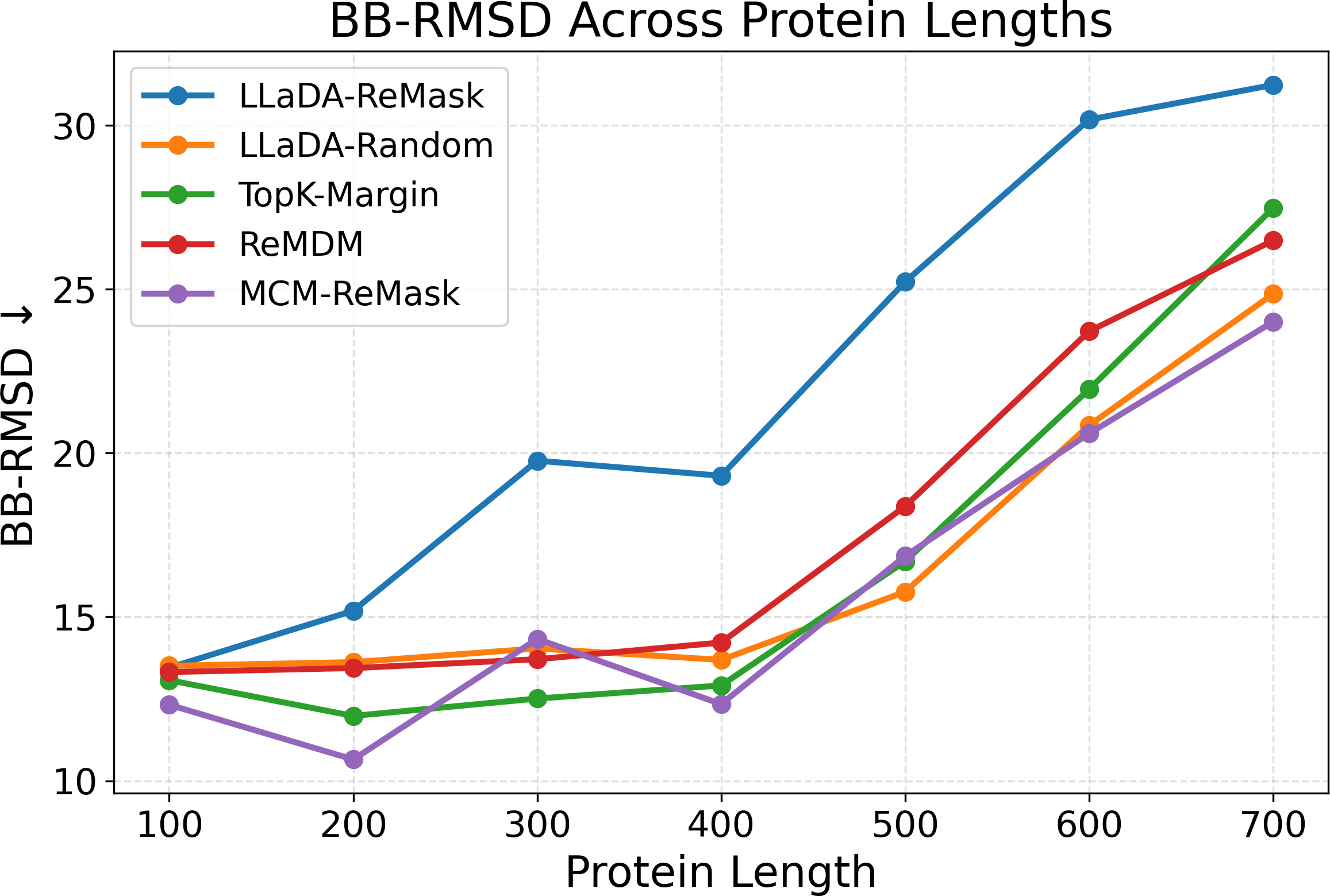}
    \caption{BB-RMSD}
    \label{fig:bb_rmsd_comparison}
\end{subfigure}
\hfill
\begin{subfigure}{0.32\textwidth}
    \centering
    \includegraphics[width=\linewidth]{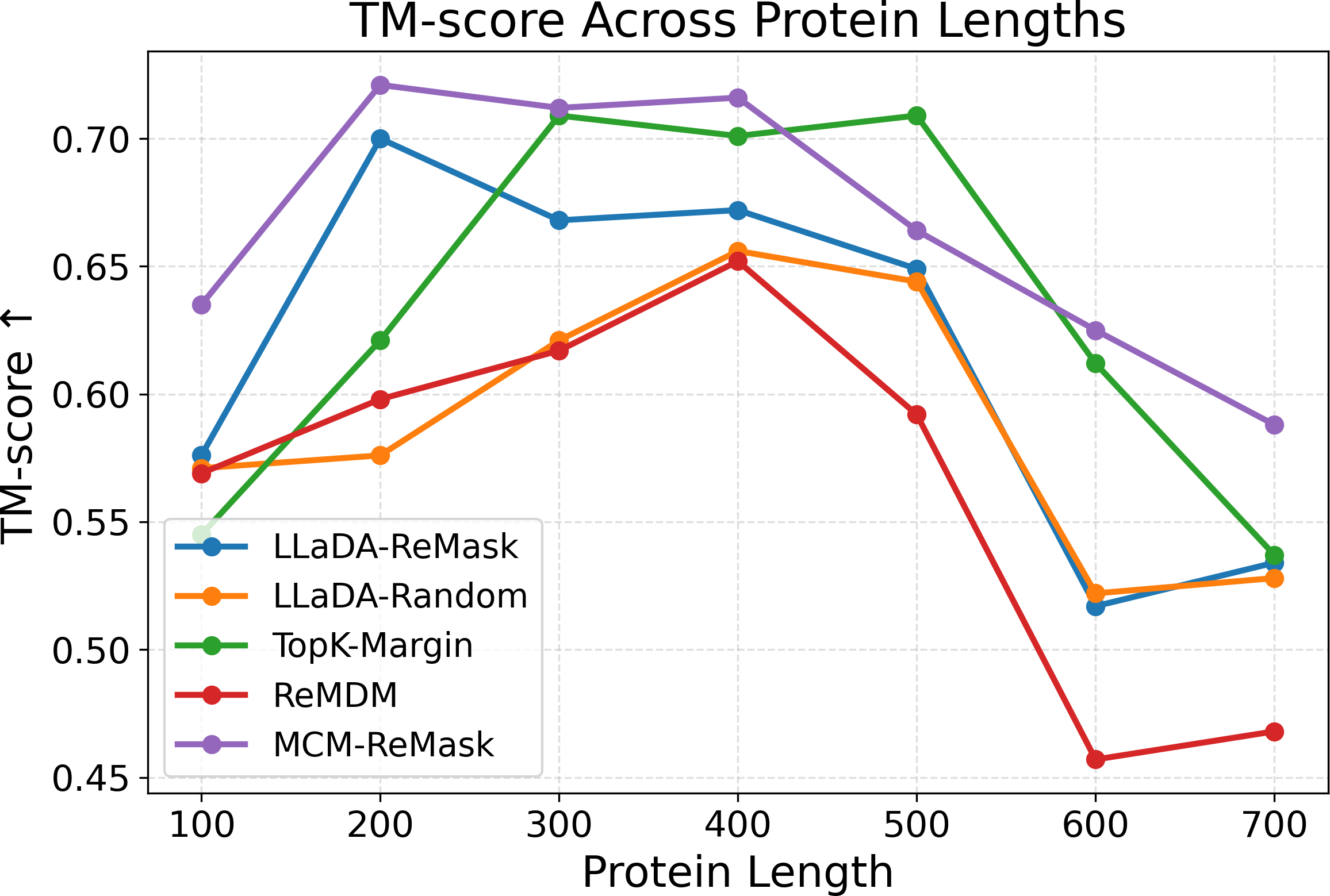}
    \caption{TM-score}
    \label{fig:tm_score_comparison}
\end{subfigure}
\hfill
\begin{subfigure}{0.32\textwidth}
    \centering
    \includegraphics[width=\linewidth]{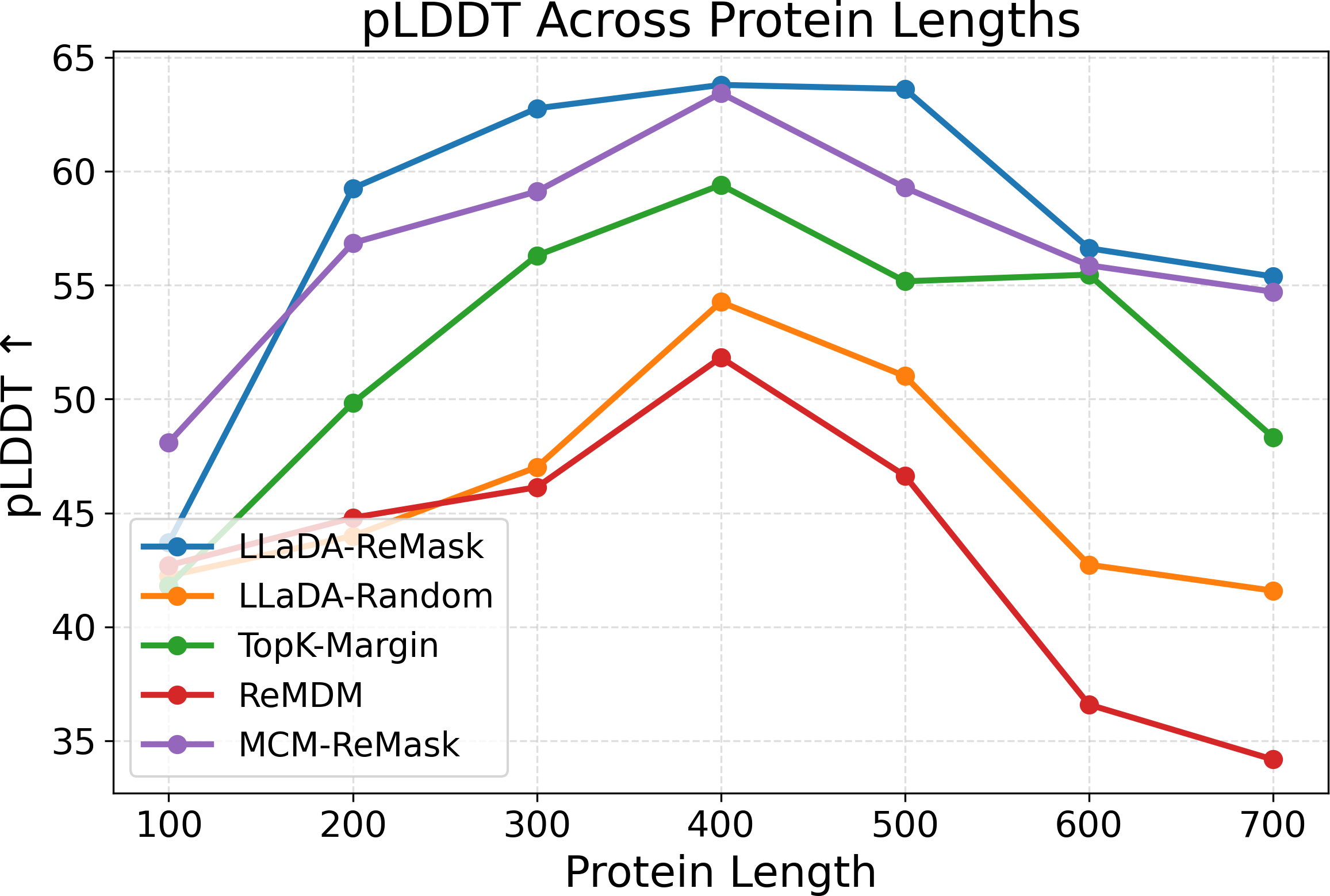}
    \caption{pLDDT}
    \label{fig:plddt_comparison}
\end{subfigure}

\caption{Comparison of different unmasking strategies across protein lengths.}
\label{fig:decoding_strategy_comparison}
\end{figure*}

We first examine whether MCM-ReMask improves sequence--structure self-consistency during discrete diffusion sampling, following the protocol described in the Section~\ref{setup}. As shown in Fig.~\ref{fig:decoding_strategy_comparison}, unmasking strategy substantially affects sequence-structure self-consistency across protein lengths. MCM-ReMask achieves the strongest TM-score over most lengths, especially from 100 to 500 residues, indicating improved global agreement between the GCP-VQVAE-decoded backbone and the ESMFold-predicted structure. It also obtains low BB-RMSD across multiple lengths and remains competitive elsewhere, suggesting better control of backbone-level inconsistency during decoding. For pLDDT, MCM-ReMask is consistently among the stronger methods, although LLaDA-ReMask gives higher confidence at some medium and long lengths; however, these gains do not always coincide with better TM-score or BB-RMSD. Overall, MCM-ReMask provides the best balance across TM-score, BB-RMSD, and pLDDT, supporting the effectiveness of confidence-margin-based verification and remasking. Full numerical results are provided in Appendix~\ref{app:unmasking_strategy_evaluation_table}. We therefore use MCM-ReMask as the default decoding strategy for ProtLiD$^2$ in subsequent experiments.

\subsection{Ligand-Conditioned Whole Protein Co-design}

We next evaluate ProtLiD$^2$ on ligand-conditioned whole-protein co-design. Existing discrete diffusion protein language models such as DPLM do not directly support small-molecule ligand conditioning, and are therefore not directly comparable in this setting. We instead compare ProtLiD$^2$ with Complexa~\cite{didi2026scaling}, a recent protein complex generation model.

The comparison is conducted on the 200-target benchmark described in Section~\ref{datac}. Following the protocol in Section~\ref{setup}, both methods are conditioned on the same ligand and target protein length for each target. Structure metrics are evaluated on all 200 targets, while AF3-Vina scores are reported on 191 targets because 9 ligands could not be converted into valid AutoDock Vina-compatible representations. Full evaluation details are provided in Appendix~\ref{app:whole_protein_eval}.

\begin{table*}[!t]
\centering
\caption{
Comparison between ProtLiD$^2$ and Complexa.
Structure metrics are evaluated on 200 targets, while AF3-Vina scores are evaluated on 191 valid targets.
}
\label{tab:ld2pcd_proteinacomplexa_comparison_with_n}

\resizebox{0.92\textwidth}{!}{
\begin{tabular}{lccccc}
\toprule
\multirow{2}{*}{Method}
& \multicolumn{2}{c}{Global RMSD $\downarrow$}
& \multirow{2}{*}{\begin{tabular}{c}TM-score $\uparrow$ \\ $(n=200)$\end{tabular}}
& \multirow{2}{*}{\begin{tabular}{c}pLDDT $\uparrow$ \\ $(n=200)$\end{tabular}}
& \multirow{2}{*}{\begin{tabular}{c}AF3-Vina $\downarrow$ \\ $(n=191)$\end{tabular}} \\
\cmidrule(lr){2-3}
& BB $(n=200)$ & CA $(n=200)$ & & & \\
\midrule
Complexa
& $\mathbf{10.35 \pm 11.24}$
& $\mathbf{10.40 \pm 11.27}$
& $0.672 \pm 0.333$
& $64.55 \pm 18.10$
& $\mathbf{-7.11 \pm 2.26}$ \\

ProtLiD$^2$
& $12.07 \pm 13.16$
& $12.13 \pm 13.15$
& $\mathbf{0.802 \pm 0.175}$
& $\mathbf{73.00 \pm 12.85}$
& $-6.82 \pm 1.75$ \\
\bottomrule
\end{tabular}
}
\end{table*}

\begin{table*}[t]
\centering
\caption{
Combined pass-rate comparison between methods.
For each criterion, we report the number of passed designs and pass rate.
Full criterion definitions are provided in Appendix~\ref{app:whole_protein_eval}.
}
\label{tab:strict_combined_pass_rate_three_methods}
\resizebox{\textwidth}{!}{
\begin{tabular}{lcccccccccc}
\toprule
\multirow{2}{*}{Method}
& \multicolumn{2}{c}{FC}
& \multicolumn{2}{c}{HCF}
& \multicolumn{2}{c}{BC-5}
& \multicolumn{2}{c}{BC-7}
& \multicolumn{2}{c}{SWPS} \\
\cmidrule(lr){2-3}
\cmidrule(lr){4-5}
\cmidrule(lr){6-7}
\cmidrule(lr){8-9}
\cmidrule(lr){10-11}
& Count & Rate
& Count & Rate
& Count & Rate
& Count & Rate
& Count & Rate \\
\midrule

Complexa
& $102/200$ & $51.00$
& $13/200$ & $6.50$
& $90/191$ & $47.12$
& $56/191$ & $29.32$
& $8/191$ & $4.19$ \\

ProtLiD$^2$
& $\mathbf{117/200}$ & $\mathbf{58.50}$
& $13/200$ & $6.50$
& $\mathbf{101/191}$ & $\mathbf{52.88}$
& $\mathbf{58/191}$ & $\mathbf{30.37}$
& $\mathbf{11/191}$ & $\mathbf{5.76}$ \\
\bottomrule
\end{tabular}
}
\end{table*}

As shown in Table~\ref{tab:ld2pcd_proteinacomplexa_comparison_with_n}, Complexa obtains lower global RMSD and slightly better AF3-Vina score, suggesting stronger coordinate agreement and docking energy in some cases. In contrast, ProtLiD$^2$ achieves substantially higher TM-score and pLDDT, indicating better global fold consistency and higher sequence foldability. This suggests that although ProtLiD$^2$ may not always minimize RMSD to the reference structure, it preserves the overall fold more reliably. We further evaluate five combined pass-rate criteria that jointly consider structural plausibility, prediction confidence, and ligand-aware docking quality: FC measures fold confidence, HCF applies a stricter fold-quality criterion, BC-5 and BC-7 additionally require AF3-Vina scores below $-5.0$ and $-7.0$, respectively, and SWPS denotes the strictest criterion. As shown in Table~\ref{tab:strict_combined_pass_rate_three_methods}, ProtLiD$^2$ improves FC from 51.00\% to 58.50\%, BC-5 from 47.12\% to 52.88\%, and BC-7 from 29.32\% to 30.37\%, while also achieving a higher SWPS pass rate. These results suggest that ProtLiD$^2$ is competitive for ligand-conditioned whole-protein design when global fold consistency, sequence-structure compatibility, and ligand-aware evaluation are considered jointly.

\subsection{Pocket Co-design}


We further evaluate ProtLiD$^2$ on ligand-binding pocket co-design, following the pocket definition and evaluation protocol in Section~\ref{setup}. Starting from the 200-target benchmark described in Section~\ref{datac}, we remove 50 multichain complexes to focus on single-chain pocket design, resulting in 150 valid targets. Vina-related metrics and pass rates are computed on 149 targets because one target could not be processed by AutoDock Vina. Full evaluation details are provided in Appendix~\ref{app:pocket_eval}.

\begin{table*}[t]
\centering
\caption{
Overall comparison of pocket-design methods.
Values are reported as mean $\pm$ standard deviation.
Lower is better for RMSD and Vina score; higher is better for TM-score and pLDDT.
}
\label{tab:pocket_design_metric_summary}
\resizebox{\textwidth}{!}{
\begin{tabular}{lccccccc}
\toprule
\multirow{2}{*}{Method}
& \multicolumn{2}{c}{Active-site RMSD $\downarrow$}
& \multicolumn{2}{c}{Global RMSD $\downarrow$}
& \multirow{2}{*}{TM-score $\uparrow$}
& \multirow{2}{*}{pLDDT $\uparrow$}
& \multirow{2}{*}{Vina $\downarrow$} \\
\cmidrule(lr){2-3}
\cmidrule(lr){4-5}
& BB & CA & BB & CA & & & \\
\midrule

FAIR
& $3.46 \pm 2.62$
& $3.37 \pm 2.69$
& $3.78 \pm 4.75$
& $3.81 \pm 4.77$
& $0.866 \pm 0.194$
& $79.83 \pm 11.31$
& $-6.94 \pm 1.74$ \\

PocketGen
& $3.40 \pm 2.54$
& $3.50 \pm 2.55$
& $3.70 \pm 4.67$
& $3.74 \pm 4.68$
& $0.869 \pm 0.192$
& $\mathbf{80.83 \pm 11.59}$
& $\mathbf{-8.84 \pm 3.80}$ \\

ProtLiD$^2$
& $\mathbf{1.97 \pm 1.69}$
& $\mathbf{2.06 \pm 1.72}$
& $\mathbf{3.63 \pm 4.45}$
& $\mathbf{3.69 \pm 4.46}$
& $\mathbf{0.915 \pm 0.127}$
& $79.17 \pm 11.49$
& $-6.93 \pm 1.48$ \\

\bottomrule
\end{tabular}
}
\end{table*}

\begin{figure*}[t]
\centering
\includegraphics[width=0.94\textwidth]{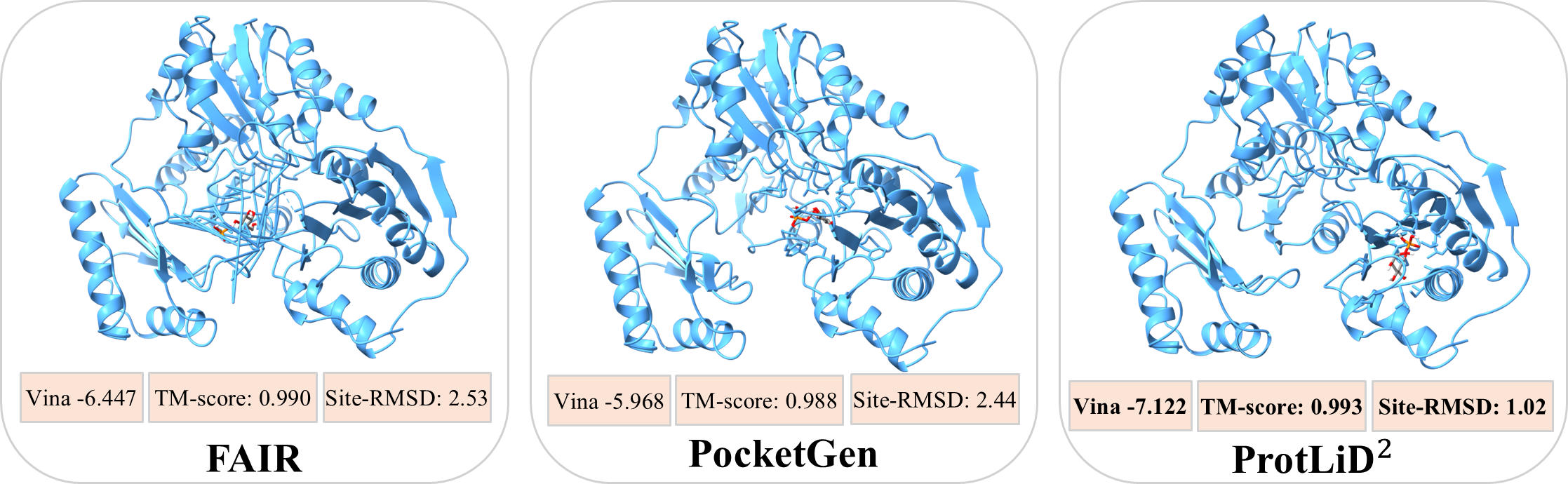}
\caption{
Qualitative pocket co-design case study on 3BKQ.
ProtLiD$^2$ improves active-site geometry with lower Site-RMSD while preserving global fold similarity and ligand compatibility.
}
\label{fig:pocket_case_study}
\end{figure*}

As shown in Table~\ref{tab:pocket_design_metric_summary}, ProtLiD$^2$ achieves the best active-site accuracy, reducing active-site BB/CA-RMSD to $1.97/2.06$, compared with $3.46/3.37$ for FAIR and $3.40/3.50$ for PocketGen. ProtLiD$^2$ also obtains the lowest global RMSD and the highest TM-score, indicating stronger global sequence--structure consistency. Although PocketGen achieves the best average Vina score and slightly higher pLDDT, ProtLiD$^2$ produces more accurate ligand-binding pocket geometry while maintaining better global fold consistency. Figure~\ref{fig:pocket_case_study} provides representative qualitative examples consistent with the aggregate results. Across the three shown targets, ProtLiD$^2$ maintains high TM-score while producing more accurate ligand-binding pocket geometry, as reflected by substantially lower active-site RMSD than FAIR and PocketGen. These cases illustrate that the improvement of ProtLiD$^2$ is not only reflected in global fold metrics, but also in local ligand-binding site reconstruction. Additional qualitative examples are provided in Appendix~\ref{fig:pocket_case_study_appendix}.

\begin{table*}[!t]
\centering
\caption{
Combined pass-rate comparison among FAIR, PocketGen, and ProtLiD$^2$.
For each criterion, we report the number of passed designs and pass rate.
Full criterion definitions are provided in Appendix~\ref{app:pocket_eval}.
}
\label{tab:combined_pass_rate_ld2pcd_pocketgen_fair}
\resizebox{\textwidth}{!}{
\begin{tabular}{lcccccccccccc}
\toprule
\multirow{2}{*}{Method}
& \multicolumn{2}{c}{FC}
& \multicolumn{2}{c}{HCF}
& \multicolumn{2}{c}{PGC}
& \multicolumn{2}{c}{BC-5}
& \multicolumn{2}{c}{BC-7}
& \multicolumn{2}{c}{SDS} \\
\cmidrule(lr){2-3}
\cmidrule(lr){4-5}
\cmidrule(lr){6-7}
\cmidrule(lr){8-9}
\cmidrule(lr){10-11}
\cmidrule(lr){12-13}
& Count & Rate
& Count & Rate
& Count & Rate
& Count & Rate
& Count & Rate
& Count & Rate \\
\midrule
FAIR
& $\mathbf{127/149}$ & $\mathbf{85.23}$
& $68/149$ & $45.64$
& $15/149$ & $10.07$
& $10/149$ & $6.71$
& $2/149$ & $1.34$
& $0/149$ & $0.00$ \\

PocketGen
& $126/149$ & $84.56$
& $\mathbf{70/149}$ & $\mathbf{46.98}$
& $30/149$ & $20.13$
& $22/148$ & $14.86$
& $9/148$ & $6.08$
& $0/148$ & $0.00$ \\

ProtLiD$^2$
& $\mathbf{127/149}$ & $\mathbf{85.23}$
& $63/149$ & $42.28$
& $\mathbf{96/149}$ & $\mathbf{64.43}$
& $\mathbf{89/149}$ & $\mathbf{59.73}$
& $\mathbf{35/149}$ & $\mathbf{23.49}$
& $\mathbf{6/149}$ & $\mathbf{4.03}$ \\

\bottomrule
\end{tabular}
}
\end{table*}


We further compare the methods using combined pass-rate criteria that jointly measure global fold confidence, active-site accuracy, and ligand-aware docking quality. FC and HCF evaluate global fold quality under standard and stricter thresholds, PGC additionally requires accurate active-site geometry, BC-5 and BC-7 further impose Vina-score thresholds of $-5.0$ and $-7.0$, and SDS denotes the strictest design-success criterion. Full definitions are provided in Appendix~\ref{app:pocket_eval}. As shown in Table~\ref{tab:combined_pass_rate_ld2pcd_pocketgen_fair}, ProtLiD$^2$ matches the best FC pass rate and substantially improves criteria involving pocket geometry and binding compatibility. In particular, ProtLiD$^2$ improves PGC from 20.13\% to 64.43\% over PocketGen, BC-5 from 14.86\% to 59.73\%, and BC-7 from 6.08\% to 23.49\%. Under the strictest SDS criterion, only ProtLiD$^2$ obtains successful designs. These results indicate that ProtLiD$^2$ is particularly effective for ligand-binding pocket co-design, where accurate local pocket geometry must be achieved together with globally plausible sequence--structure generation.







\section{Discussion}

ProtLiD$^2$ demonstrates that ligand conditioning can be integrated into masked discrete diffusion over unified sequence-structure tokens. The proposed MCM-ReMask decoding first improves sequence-structure self-consistency through lightweight inference-time self-correction. Building on this decoding strategy, ProtLiD$^2$ improves whole-protein TM-score and pLDDT over Complexa, and substantially reduces active-site RMSD while increasing ligand-aware pass rates over FAIR and PocketGen in pocket co-design. These results suggest that ProtLiD$^2$ combines robust token-space generation with geometry-aware ligand conditioning for functional protein co-design.

Several limitations and broader-impact considerations remain. ProtLiD$^2$ relies on a frozen backbone tokenizer, lacks explicit full-atom side-chain and ligand-flexibility modeling, and is evaluated mainly with computational proxies such as ESMFold, AlphaFold3, and AutoDock Vina; thus, generated proteins require experimental validation. While ProtLiD$^2$ may accelerate ligand-aware protein and enzyme design, generative protein design also carries dual-use risks and should be accompanied by expert review, biosafety screening, and safeguards for future releases. To support reproducibility, we will release the inference and evaluation code after refactoring, together with the training and validation datasets and reproduction instructions. Future work will explore improved tokenization, full-atom refinement, stronger ligand-aware objectives, and experimental validation.

\bibliographystyle{unsrt}
\bibliography{reference}

\newpage
\appendix
\section{Appendices}

\appendix

\setcounter{section}{0}
\renewcommand{\thesection}{A\arabic{section}}

\setcounter{table}{0}
\renewcommand{\thetable}{A\arabic{table}}

\setcounter{figure}{0}
\renewcommand{\thefigure}{A\arabic{figure}}

\section{Dataset Processing Details}
\label{app:dataset_details}

We integrated ligand-protein complexes from Protenix~\cite{2026.02.05.703733}, PLINDER~\cite{anonymous2024plinder}, CrossDock~\cite{doi:10.1021/acs.jcim.0c00411}, HiQBind~\cite{D4DD00357H}, and AlphaFill-derived complexes~\cite{Hekkelman2023}. Ligand-protein complexes were first extracted from the Protenix training set. Complexes from the remaining sources were processed using the AlphaFold3 data-processing pipeline~\cite{Abramson2024}, which provides unified parsing and cleanup of biomolecular complexes, including resolving alternative atom locations, removing waters, normalizing residue names, and expanding biological assemblies.

For each candidate complex, we extracted a single protein chain and its associated ligand. The protein component was represented by the amino-acid sequence and residue-wise coordinates of the four main-chain atoms N, C$_\alpha$, C, and O. The ligand component was represented by atom coordinates, atom types, ligand identifiers, and SMILES strings. Complexes were retained only when both protein and ligand could be parsed successfully, the protein sequence was consistent with the coordinate-derived sequence, a valid ligand SMILES was available, and at least one ligand-contacting residue was identified within a 6.0~\AA{} protein-ligand distance cutoff.

We removed proteins longer than 1000 residues and ligands containing more than 100 atoms. Complexes with severe protein-ligand steric clashes were discarded. A clash was defined by either an absolute interatomic distance below 0.8~\AA{} or a van der Waals overlap criterion with a source-dependent tolerance of 0.6~\AA{}. For AlphaFill-derived complexes, we retained only complexes with mean predicted confidence greater than 80.

After filtering and source-specific selection, the dataset contained 395,142 complexes from Protenix, 116,134 from CrossDock, 27,150 from HiQBind, and 115,610 from PLINDER. For AlphaFill-derived data, AlphaFold Database protein models are enriched with small molecules, cofactors, and ions transplanted from homologous experimentally determined structures. After filtering, this source yielded 5,281,501 processed complexes, from which we randomly sampled 587,136 complexes to balance the training data across sources. The final merged dataset contained 1,125,038 ligand-protein complexes.

To prevent training-test leakage, we compared training protein sequences against PLINDER test proteins using MMseqs2~\cite{Steinegger2017} and removed training examples with sequence identity $\geq 30\%$ to any benchmark protein. This de-overlap procedure produced the final training set of 1,026,766 ligand-protein complexes.

\section{Model Details}
\label{app:model_details}

\subsection{Coordinate Canonicalization}

Given protein backbone coordinates $\mathbf{X}_{\mathrm{p}}\in\mathbb{R}^{L\times4\times3}$ and ligand heavy-atom coordinates $\mathbf{X}_{\ell}\in\mathbb{R}^{M\times3}$, we canonicalize each protein-ligand complex to reduce global translational and rotational variation. We first translate the complex by the protein C$\alpha$ centroid,
\[
\mathbf{c}_{\mathrm{p}}
=
\frac{1}{L}
\sum_{i=1}^{L}
\mathbf{X}_{\mathrm{p}}^{(i,\mathrm{C}\alpha)} .
\]
This yields centered coordinates
$\widetilde{\mathbf{X}}_{\mathrm{p}}=\mathbf{X}_{\mathrm{p}}-\mathbf{c}_{\mathrm{p}}$ and
$\widetilde{\mathbf{X}}_{\ell}=\mathbf{X}_{\ell}-\mathbf{c}_{\mathrm{p}}$.
We then compute a deterministic ligand-guided PCA rotation matrix
$\mathbf{R}\in\mathbb{R}^{3\times3}$ from the centered ligand coordinates
$\widetilde{\mathbf{X}}_{\ell}$ and apply the same rigid rotation to both the protein and ligand:
\[
\mathbf{X}'_{\mathrm{p}}
=
\widetilde{\mathbf{X}}_{\mathrm{p}}\mathbf{R},
\qquad
\mathbf{X}'_{\ell}
=
\widetilde{\mathbf{X}}_{\ell}\mathbf{R}.
\]
This transformation preserves the relative protein-ligand geometry while providing a consistent coordinate frame for ligand-conditioned generation.

\subsection{Ligand Embedding Module}

\begin{figure}[t]
    \centering
    \includegraphics[width=\linewidth]{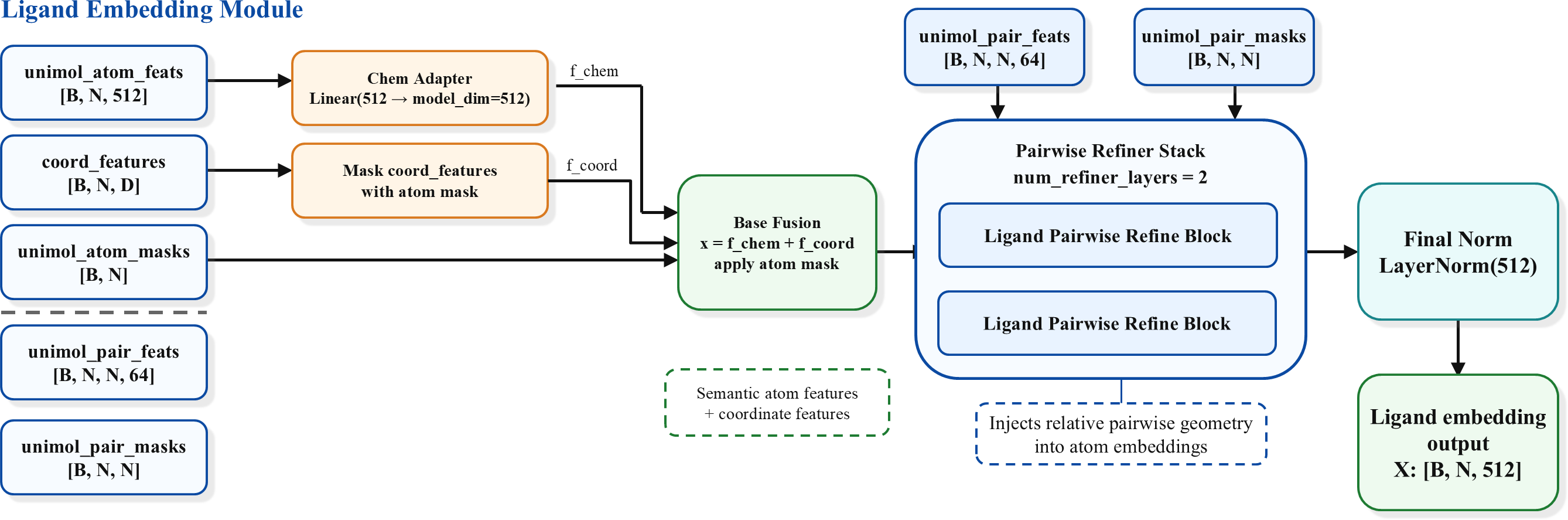}
    \caption{
    Ligand embedding module. Uni-Mol atom features and Fourier coordinate embeddings are fused to obtain initial ligand atom representations. Uni-Mol pair features and pair masks are then used by stacked pairwise refinement blocks to inject relative atom-atom geometry into ligand embeddings. The final normalized ligand embedding is used as the ligand memory for cross-attention.
    }
    \label{fig:ligand_embedding_module}
\end{figure}

The ligand embedding module is shown in Fig.~\ref{fig:ligand_embedding_module}. For each ligand with $M$ atoms, the input consists of Uni-Mol atom features, ligand coordinate features, atom masks, Uni-Mol pair features, and pair masks. Let
$\mathbf{U}\in\mathbb{R}^{B\times M\times d_{\mathrm{atom}}}$ denote Uni-Mol atom features and
$\mathbf{P}\in\mathbb{R}^{B\times M\times M\times d_{\mathrm{pair}}}$ denote Uni-Mol pair features. In our implementation, the Uni-Mol atom features have dimension $d_{\mathrm{atom}}=512$, while the pair features have dimension $d_{\mathrm{pair}}=64$.

The Uni-Mol atom features are first projected into the model hidden dimension by a chemical adapter,
\[
\mathbf{F}_{\mathrm{chem}}
=
\mathbf{U}\mathbf{W}_{\mathrm{chem}}.
\]
In parallel, the ligand coordinates are encoded by Fourier coordinate embeddings, producing coordinate features
$\mathbf{F}_{\mathrm{coord}}\in\mathbb{R}^{B\times M\times d}$.
Invalid atoms are removed by the atom mask, and the initial ligand representation is obtained by feature fusion:
\[
\mathbf{H}^{(0)}_{\ell}
=
\left(
\mathbf{F}_{\mathrm{chem}}
+
\mathbf{F}_{\mathrm{coord}}
\right)
\odot
\mathbf{m}_{\ell},
\]
where $\mathbf{m}_{\ell}\in\{0,1\}^{B\times M}$ is the ligand atom mask, broadcast along the hidden dimension.

To further refine ligand atom embeddings, we use a stack of pairwise-aware refinement blocks. These blocks inject Uni-Mol pairwise atom-atom information into atom representations through pair-biased self-attention. Specifically, the pair representation is symmetrized and projected into a head-wise attention bias:
\[
\mathbf{B}_{\mathrm{pair}}
=
\mathrm{Proj}_{\mathrm{pair}}
\left(
\frac{\mathbf{P}+\mathbf{P}^{\top}}{2}
\right).
\]
The ligand self-attention is then computed as
\[
\mathrm{Attn}_{\ell}
=
\mathrm{Softmax}
\left(
\frac{\mathbf{Q}\mathbf{K}^{\top}}{\sqrt{d_h}}
+
\alpha_{\mathrm{pair}}\mathbf{B}_{\mathrm{pair}}
+
\mathbf{B}_{\mathrm{mask}}
\right)
\mathbf{V},
\]
where $\alpha_{\mathrm{pair}}$ is a learnable pair-bias scale, and $\mathbf{B}_{\mathrm{mask}}$ is the attention mask derived from the ligand atom and pair masks. After $K$ pairwise refinement layers, the final ligand memory is obtained by layer normalization:
\[
\mathbf{M}_{\ell}
=
\mathrm{LayerNorm}
\left(
\mathbf{H}^{(K)}_{\ell}
\right),
\qquad
\mathbf{M}_{\ell}\in\mathbb{R}^{B\times M\times d}.
\]
This ligand memory contains both semantic atom-level chemical information and coordinate-aware geometric information, and is used as the conditioning memory in the protein denoising Transformer.

\subsection{Geometry-Aware Ligand Cross-Attention}

\begin{figure}[t]
    \centering
    \includegraphics[width=\linewidth]{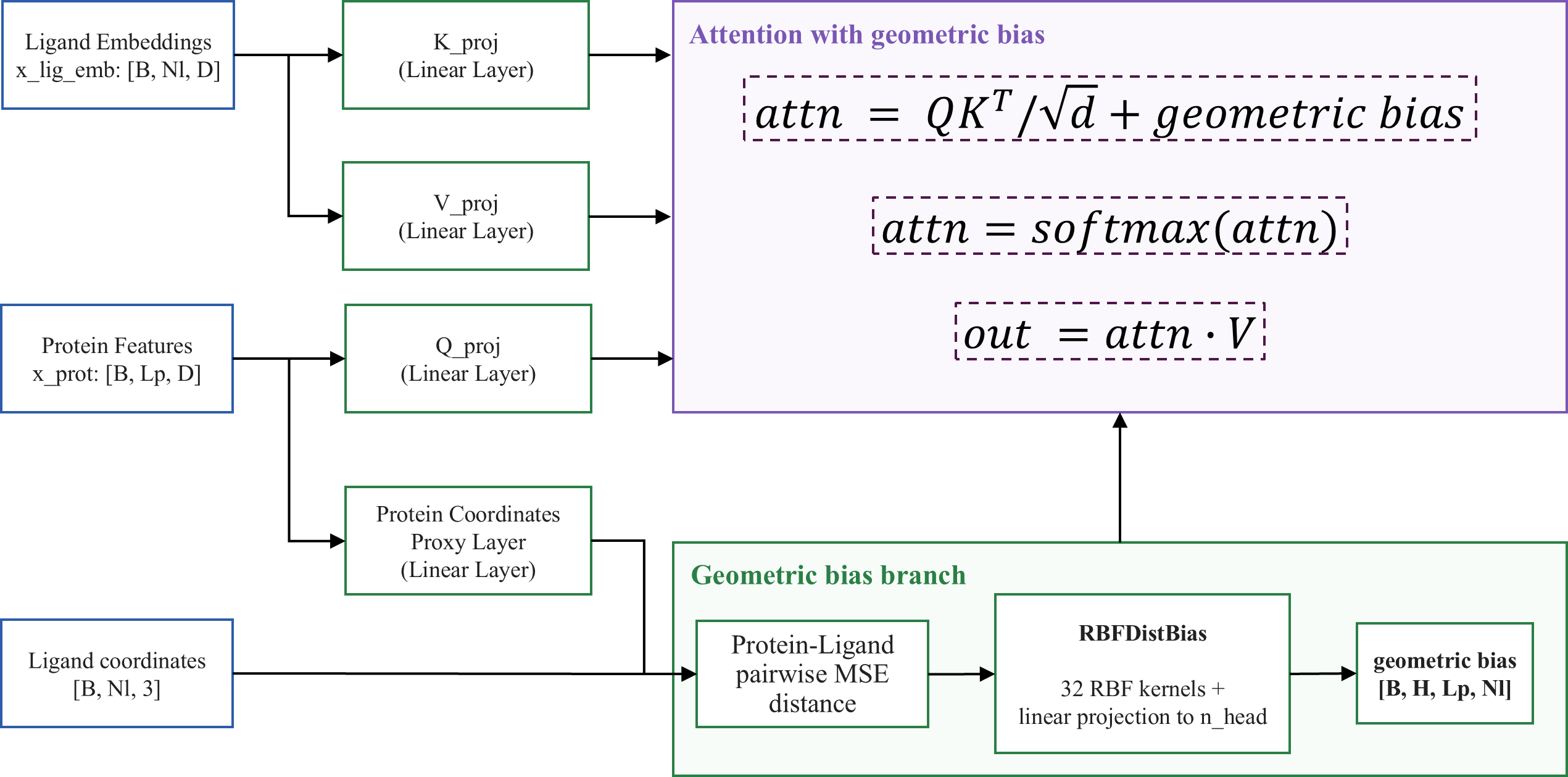}
    \caption{
    Geometry-aware ligand cross-attention module. Protein hidden states provide queries, while ligand embeddings provide keys and values. A separate geometric branch predicts protein-token proxy coordinates, computes protein-ligand pairwise distances against ligand atom coordinates, and converts them into a head-wise geometric bias. The final cross-attention logits combine content-based attention and geometric bias.
    }
    \label{fig:geometry_cross_attention}
\end{figure}

The geometry-aware ligand cross-attention module is illustrated in Fig.~\ref{fig:geometry_cross_attention}. Given protein hidden states
$\mathbf{H}\in\mathbb{R}^{B\times T\times d}$ and ligand memory
$\mathbf{M}_{\ell}\in\mathbb{R}^{B\times M\times d}$, the protein hidden states are used as queries, while ligand embeddings provide keys and values:
\[
\mathbf{Q}
=
\mathrm{RMSNorm}(\mathbf{H}\mathbf{W}_{Q}),
\qquad
\mathbf{K}_{\ell}
=
\mathrm{RMSNorm}(\mathbf{M}_{\ell}\mathbf{W}_{K}),
\qquad
\mathbf{V}_{\ell}
=
\mathbf{M}_{\ell}\mathbf{W}_{V}.
\]
The standard content-based cross-attention logits are
\[
\mathbf{A}_{\mathrm{content}}
=
\frac{\mathbf{Q}\mathbf{K}_{\ell}^{\top}}{\sqrt{d_h}}.
\]

To make ligand conditioning sensitive to protein-ligand geometry, we add a geometric bias branch. Each protein token hidden state is first mapped to a learned proxy coordinate:
\[
\widehat{\mathbf{R}}
=
\rho\cdot\tanh(\mathbf{H}\mathbf{W}_{r}),
\qquad
\widehat{\mathbf{R}}\in\mathbb{R}^{B\times T\times3},
\]
where $\rho=20$ constrains the proxy coordinates to a bounded spatial range. We then compute pairwise distances between the predicted protein-token proxy coordinates and the ligand atom coordinates:
\[
d_{ij}
=
\left\|
\widehat{\mathbf{r}}_i
-
\mathbf{x}^{\ell\prime}_j
\right\|_2,
\qquad
\mathbf{D}\in\mathbb{R}^{B\times T\times M}.
\]
The distance matrix is expanded with radial basis functions and projected into a head-wise geometric bias:
\[
\mathbf{B}_{\mathrm{geom}}
=
\mathrm{Proj}_{\mathrm{rbf}}
\left(
\mathrm{RBF}(\mathbf{D})
\right),
\qquad
\mathbf{B}_{\mathrm{geom}}
\in
\mathbb{R}^{B\times H\times T\times M}.
\]
The final attention logits combine content-based similarity and geometric bias:
\[
\mathbf{A}
=
\frac{\mathbf{Q}\mathbf{K}_{\ell}^{\top}}{\sqrt{d_h}}
+
\mathbf{B}_{\mathrm{geom}}.
\]
The ligand-conditioned cross-attention output is then
\[
\mathrm{CrossAttn}_{\ell}
=
\mathrm{Softmax}(\mathbf{A})\mathbf{V}_{\ell}.
\]
The output is projected back to the model dimension and added to the protein hidden states through a residual connection. This design allows protein sequence and structure tokens to attend to ligand atoms using both chemical compatibility and learned spatial proximity.

\section{Experiment}

\subsection{Unmasking Strategy Evaluation for Protein Co-design}
\label{app:unmasking_strategy_evaluation}

We provide the complete unmasking strategy comparison across protein lengths from 100 to 700 residues. For each decoding strategy and target length, 100 protein sequence-structure pairs were generated and evaluated by comparing the GCP-VQVAE-decoded backbone structure with the ESMFold-predicted structure from the generated sequence. The results are reported as mean $\pm$ standard deviation for CA-RMSD, BB-RMSD, TM-score, and pLDDT, where lower RMSD and higher TM-score/pLDDT indicate better sequence-structure self-consistency.

\begin{table*}[!t]
\centering
\caption{Comparison of different decoding methods across protein lengths.}
\label{app:unmasking_strategy_evaluation_table}
\resizebox{\textwidth}{!}{
\begin{tabular}{llcccc}
\hline
Method & Length & CA-RMSD $\downarrow$ & BB-RMSD $\downarrow$ & TM-score $\uparrow$ & pLDDT $\uparrow$ \\
\hline
LLaDA-ReMask & 100 & $13.55 \pm 5.85$ & $13.45 \pm 5.86$ & $0.576 \pm 0.164$ & $43.69 \pm 13.40$ \\
LLaDA-Random & 100 & $13.62 \pm 4.49$ & $13.51 \pm 4.49$ & $0.571 \pm 0.149$ & $42.24 \pm 10.44$ \\
TopK-Margin  & 100 & $13.18 \pm 4.36$ & $13.07 \pm 4.37$ & $0.545 \pm 0.152$ & $41.80 \pm 10.41$ \\
ReMDM        & 100 & $13.43 \pm 3.40$ & $13.32 \pm 3.41$ & $0.569 \pm 0.135$ & $42.70 \pm 10.17$ \\
MCM-ReMask   & 100 & $\mathbf{12.43 \pm 5.24}$ & $\mathbf{12.32 \pm 5.25}$ & $\mathbf{0.635 \pm 0.144}$ & $\mathbf{48.09 \pm 12.54}$ \\
\hline
LLaDA-ReMask & 200 & $15.28 \pm 14.12$ & $15.19 \pm 14.14$ & $0.700 \pm 0.214$ & $\mathbf{59.25 \pm 18.53}$ \\
LLaDA-Random & 200 & $13.72 \pm 6.05$  & $13.62 \pm 6.04$  & $0.576 \pm 0.183$ & $43.99 \pm 17.18$ \\
TopK-Margin  & 200 & $12.08 \pm 6.77$  & $11.98 \pm 6.75$  & $0.621 \pm 0.217$ & $49.83 \pm 19.85$ \\
ReMDM        & 200 & $13.54 \pm 6.26$  & $13.44 \pm 6.25$  & $0.598 \pm 0.199$ & $44.79 \pm 18.50$ \\
MCM-ReMask   & 200 & $\mathbf{10.74 \pm 6.31}$ & $\mathbf{10.65 \pm 6.31}$ & $\mathbf{0.721 \pm 0.169}$ & $56.85 \pm 17.16$ \\
\hline
LLaDA-ReMask & 300 & $19.84 \pm 22.05$ & $19.77 \pm 22.06$ & $0.668 \pm 0.266$ & $\mathbf{62.76 \pm 17.46}$ \\
LLaDA-Random & 300 & $14.13 \pm 8.00$  & $14.04 \pm 8.00$  & $0.621 \pm 0.199$ & $47.01 \pm 16.69$ \\
TopK-Margin  & 300 & $\mathbf{12.60 \pm 10.85}$ & $\mathbf{12.51 \pm 10.85}$ & $0.709 \pm 0.223$ & $56.30 \pm 18.61$ \\
ReMDM        & 300 & $13.80 \pm 8.06$  & $13.71 \pm 8.06$  & $0.617 \pm 0.216$ & $46.13 \pm 17.96$ \\
MCM-ReMask   & 300 & $14.41 \pm 14.32$ & $14.33 \pm 14.33$ & $\mathbf{0.712 \pm 0.183}$ & $59.11 \pm 14.69$ \\
\hline
LLaDA-ReMask & 400 & $19.37 \pm 17.71$ & $19.31 \pm 17.71$ & $0.672 \pm 0.254$ & $\mathbf{63.79 \pm 16.79}$ \\
LLaDA-Random & 400 & $13.77 \pm 8.17$  & $13.69 \pm 8.17$  & $0.656 \pm 0.223$ & $54.26 \pm 18.82$ \\
TopK-Margin  & 400 & $12.98 \pm 9.89$  & $12.91 \pm 9.90$  & $0.701 \pm 0.235$ & $59.40 \pm 16.92$ \\
ReMDM        & 400 & $14.30 \pm 8.09$  & $14.22 \pm 8.08$  & $0.652 \pm 0.217$ & $51.82 \pm 19.24$ \\
MCM-ReMask   & 400 & $\mathbf{12.42 \pm 10.54}$ & $\mathbf{12.34 \pm 10.55}$ & $\mathbf{0.716 \pm 0.222}$ & $63.43 \pm 16.19$ \\
\hline
LLaDA-ReMask & 500 & $25.30 \pm 20.11$ & $25.23 \pm 20.13$ & $0.649 \pm 0.277$ & $\mathbf{63.61 \pm 15.28}$ \\
LLaDA-Random & 500 & $\mathbf{15.85 \pm 8.81}$ & $\mathbf{15.77 \pm 8.82}$ & $0.644 \pm 0.208$ & $51.01 \pm 20.75$ \\
TopK-Margin  & 500 & $16.78 \pm 12.60$ & $16.70 \pm 12.61$ & $\mathbf{0.709 \pm 0.236}$ & $55.17 \pm 20.00$ \\
ReMDM        & 500 & $18.46 \pm 9.15$  & $18.38 \pm 9.15$  & $0.592 \pm 0.221$ & $46.64 \pm 20.12$ \\
MCM-ReMask   & 500 & $16.93 \pm 11.45$ & $16.86 \pm 11.46$ & $0.664 \pm 0.240$ & $59.28 \pm 16.39$ \\
\hline
LLaDA-ReMask & 600 & $30.25 \pm 18.73$ & $30.18 \pm 18.74$ & $0.517 \pm 0.273$ & $\mathbf{56.62 \pm 18.57}$ \\
LLaDA-Random & 600 & $20.93 \pm 7.71$  & $20.85 \pm 7.71$  & $0.522 \pm 0.195$ & $42.73 \pm 17.23$ \\
TopK-Margin  & 600 & $22.02 \pm 12.49$ & $21.95 \pm 12.49$ & $0.612 \pm 0.253$ & $55.46 \pm 19.11$ \\
ReMDM        & 600 & $23.79 \pm 7.01$  & $23.72 \pm 7.01$  & $0.457 \pm 0.169$ & $36.59 \pm 15.64$ \\
MCM-ReMask   & 600 & $\mathbf{20.66 \pm 11.73}$ & $\mathbf{20.60 \pm 11.73}$ & $\mathbf{0.625 \pm 0.247}$ & $55.87 \pm 16.75$ \\
\hline
LLaDA-ReMask & 700 & $31.31 \pm 12.81$ & $31.24 \pm 12.81$ & $0.534 \pm 0.247$ & $\mathbf{55.38 \pm 15.75}$ \\
LLaDA-Random & 700 & $24.93 \pm 5.81$  & $24.86 \pm 5.81$  & $0.528 \pm 0.177$ & $41.59 \pm 13.84$ \\
TopK-Margin  & 700 & $27.54 \pm 10.64$ & $27.48 \pm 10.64$ & $0.537 \pm 0.230$ & $48.33 \pm 16.03$ \\
ReMDM        & 700 & $26.56 \pm 4.55$  & $26.49 \pm 4.55$  & $0.468 \pm 0.161$ & $34.18 \pm 11.92$ \\
MCM-ReMask   & 700 & $\mathbf{24.07 \pm 11.56}$ & $\mathbf{24.01 \pm 11.56}$ & $\mathbf{0.588 \pm 0.259}$ & $54.71 \pm 17.87$ \\
\hline
\end{tabular}
}
\end{table*}

\subsection{Ligand-Conditioned Whole Protein Co-design}
\label{app:whole_protein_eval}

For each benchmark target, we use the native ligand and the length of the corresponding ligand-binding protein from the PDB complex as the input condition. Each method generates 10 candidate protein designs for the same ligand and target length. We evaluate sequence-structure self-consistency by folding each generated amino-acid sequence with ESMFold and comparing the ESMFold-predicted structure with the model-generated protein structure. We report backbone RMSD, C$\alpha$ RMSD, TM-score, and pLDDT. For each target, the candidate with the highest ESMFold pLDDT is selected as the representative design for downstream ligand-aware evaluation.

To assess ligand compatibility, we use AlphaFold3 to predict the complex structure between the selected designed protein and the input ligand. Based on the AF3-predicted protein-ligand complex, we center the docking box on the ligand and compute the AF3 Vina score using AutoDock Vina.

Table~\ref{tab:whole_protein_pass_rate_criteria} defines the combined pass-rate criteria used for ligand-conditioned whole-protein co-design evaluation. These criteria progressively combine global fold similarity, model confidence, backbone-level structural agreement, and ligand-aware docking quality. FC and HCF assess whether a generated protein forms a confident and globally consistent fold, BC-5 and BC-7 further require favorable AF3-Vina scores under two docking thresholds, and SWPS represents the strictest criterion by requiring high fold confidence, low backbone RMSD, and strong predicted ligand binding simultaneously.

Among the 200 benchmark targets, AF3-Vina scores were obtained for 191 targets. The remaining 9 targets were excluded from Vina-score analysis because their ligands could not be converted into valid AutoDock Vina-compatible representations. These failures were mainly caused by RDKit sanitization errors from invalid valence assignments after ligand format conversion, or unsupported AutoDock atom types, such as Au or B, in the generated PDBQT files. Since these errors occurred during ligand preparation and PDBQT parsing, Vina scoring could not be performed for these cases. Therefore, Vina-based metrics are reported on the 191 successfully processed targets, while structure-based metrics are reported on the full target set when available.

\begin{table}[t]
\centering
\caption{
Definitions of the combined pass-rate criteria used for ligand-conditioned whole-protein co-design.
}
\label{tab:whole_protein_pass_rate_criteria}
\resizebox{\linewidth}{!}{
\begin{tabular}{ll}
\toprule
Shortcut & Criterion \\
\midrule
FC
& $\mathrm{TM}_{BB} > 0.7 \ \land\ \mathrm{pLDDT} > 70$ \\

HCF
& $\mathrm{TM}_{BB} > 0.85 \ \land\ \mathrm{pLDDT} > 85 \ \land\ \mathrm{BB\mbox{-}RMSD} < 2.0$ \\

BC-5
& $\mathrm{TM}_{BB} > 0.7 \ \land\ \mathrm{pLDDT} > 70 \ \land\ \mathrm{AF3\mbox{-}Vina} \leq -5.0$ \\

BC-7
& $\mathrm{TM}_{BB} > 0.7 \ \land\ \mathrm{pLDDT} > 70 \ \land\ \mathrm{AF3\mbox{-}Vina} \leq -7.0$ \\

SWPS
& $\mathrm{TM}_{BB} > 0.85 \ \land\ \mathrm{pLDDT} > 85 \ \land\ \mathrm{BB\mbox{-}RMSD} < 2.0 \ \land\ \mathrm{AF3\mbox{-}Vina} \leq -7.0$ \\
\bottomrule
\end{tabular}
}
\end{table}

\subsection{Pocket Co-design}
\label{app:pocket_eval}

\label{app:pocket_eval}

The pocket co-design benchmark is constructed from the 200-target benchmark dataset described in Section~\ref{datac}. Since this evaluation focuses on single-chain ligand-binding pocket design, we exclude 50 multichain complexes, resulting in 150 valid pocket-design targets. For each target, ligand-contacting active-site residues are defined as protein residues with any heavy atom within 6.0~\AA{} of any ligand heavy atom in the native protein-ligand complex. Each method is then tasked with redesigning these pocket residues while keeping the remaining protein context fixed.

\begin{table}[t]
\centering
\caption{
Definitions of the combined pass-rate criteria used for pocket co-design.
}
\label{tab:combined_pass_rate_criteria_ld2pcd_pocketgen_fair}
\resizebox{\linewidth}{!}{
\begin{tabular}{ll}
\toprule
Shortcut & Criterion \\
\midrule
FC
& $\mathrm{TM}_{BB} > 0.7 \ \land\ \mathrm{pLDDT} > 70$ \\

HCF
& $\mathrm{TM}_{BB} > 0.8 \ \land\ \mathrm{pLDDT} > 80 \ \land\ \mathrm{BB\mbox{-}RMSD} < 2.0$ \\

PGC
& $\mathrm{AS\mbox{-}local\ BB\mbox{-}RMSD} < 2.0 \ \land\ \mathrm{TM}_{BB} > 0.7 \ \land\ \mathrm{pLDDT} > 70$ \\

BC-5
& $\mathrm{AS\mbox{-}local\ BB\mbox{-}RMSD} < 2.0 \ \land\ \mathrm{TM}_{BB} > 0.7 \ \land\ \mathrm{pLDDT} > 70 \ \land\ \mathrm{Vina} \leq -5.0$ \\

BC-7
& $\mathrm{AS\mbox{-}local\ BB\mbox{-}RMSD} < 2.0 \ \land\ \mathrm{TM}_{BB} > 0.7 \ \land\ \mathrm{pLDDT} > 70 \ \land\ \mathrm{Vina} \leq -7.0$ \\

SDS
& $\mathrm{AS\mbox{-}local\ BB\mbox{-}RMSD} < 1.0 \ \land\ \mathrm{TM}_{BB} > 0.8 \ \land\ \mathrm{pLDDT} > 80 \ \land\ \mathrm{Vina} \leq -7.0$ \\
\bottomrule
\end{tabular}
}
\end{table}

For each target, each method generates 10 candidate pocket designs under the same ligand and structural context. We fold each generated amino-acid sequence using ESMFold and compare the ESMFold-predicted structure with the model-generated structure. We report both global and active-site RMSD using backbone atoms and C$\alpha$ atoms, as well as TM-score and pLDDT. Active-site RMSD is computed over ligand-contacting residues, while global RMSD and TM-score are computed over the full protein chain. For each target, the candidate with the highest ESMFold pLDDT is selected as the representative design for ligand-aware evaluation.

\begin{figure*}[t]
\centering
\includegraphics[width=\textwidth]{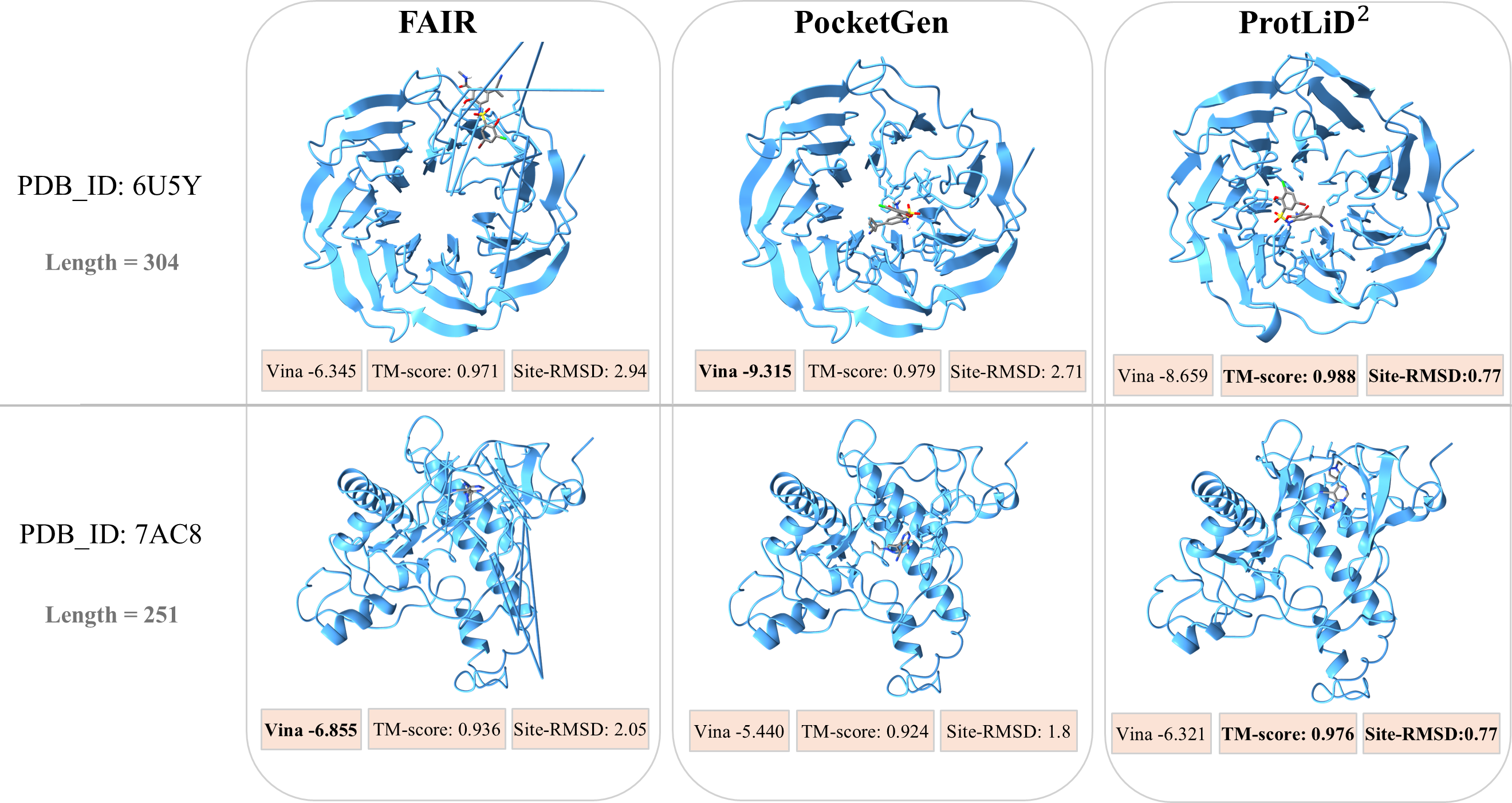}
\caption{
Additional qualitative pocket co-design case studies on 6U5Y and 7AC8.
Compared with FAIR and PocketGen, ProtLiD$^2$ achieves lower Site-RMSD while maintaining high TM-score, illustrating improved local pocket geometry across multiple targets.
}
\label{fig:pocket_case_study_appendix}
\end{figure*}

To evaluate ligand-binding plausibility, we compute AutoDock Vina scores using a ligand-centered docking box. Among the 150 valid pocket-design targets, Vina scoring was successfully performed for 149 targets. One target was excluded from Vina-based evaluation because its ligand could not be converted into a valid AutoDock Vina-compatible representation. Therefore, structure-based metrics are reported on 150 targets, while Vina-related pass-rate criteria are evaluated on the targets with valid Vina scores.

Figure~\ref{fig:pocket_case_study_appendix} shows additional qualitative pocket co-design examples on 6U5Y and 7AC8. In both cases, ProtLiD$^2$ maintains high global fold similarity while achieving the lowest Site-RMSD among the compared methods, indicating more accurate ligand-binding pocket geometry. These visual examples are consistent with the aggregate results, where ProtLiD$^2$ improves pocket accuracy and ligand-aware pass rates over FAIR and PocketGen.

Table~\ref{tab:combined_pass_rate_criteria_ld2pcd_pocketgen_fair} defines the combined pass-rate criteria used for pocket co-design evaluation. These criteria progressively combine global fold confidence, active-site geometric accuracy, and ligand-aware docking quality. FC and HCF measure global fold confidence under standard and stricter thresholds, PGC additionally requires accurate active-site geometry, BC-5 and BC-7 further impose ligand-aware docking-score thresholds, and SDS denotes the strictest success criterion by requiring accurate active-site geometry, high fold confidence, and favorable predicted ligand binding simultaneously.


\end{document}